\documentclass[twoside,twocolumn,english]{revtex4-1}
\usepackage[T1]{fontenc}
\usepackage[latin9]{inputenc}
\usepackage{color}
\usepackage{amsmath}
\usepackage{amssymb}
\usepackage{graphicx}
\usepackage{wasysym}
\usepackage{esint}

\makeatletter

\usepackage{soul}
\usepackage{color}
\usepackage{babel}

\makeatother

\usepackage{babel}
\begin{document}
\title{Analysis of landscape hierarchy during coarsening
and aging in Ising spin glasses}
\author{Stefan Boettcher and Mahajabin Rahman}
\affiliation{Department of Physics, Emory University, Atlanta, GA 30322, USA}
\begin{abstract}
We use record dynamics (RD), a coarse-grained description of the ubiquitous
relaxation phenomenology known as \textquotedbl aging\textquotedbl ,
as a diagnostic tool to find universal features that distinguish between
the energy landscapes of Ising spin models and the ferromagnet. According
to RD, a non-equilibrium system after a quench relies on fluctuations
that randomly generate a sequence of irreversible record-sized events
(quakes or avalanches) that allow the system to escape ever-higher
barriers of meta-stable states within a complex, hierarchical energy
landscape. Once these record events allow the system to overcome such
barriers, the system relaxes by tumbling into the following meta-stable
state that is marginally more stable. Within this framework, a clear
distinction can be drawn between the coarsening dynamics of an Ising
ferromagnet and the aging of the spin glass, which are often put in
the same category. To that end, we interpolate between the spin glass
and ferromagnet by varying the admixture $p$ of ferromagnetic over
anti-ferromagnetic bonds from the glassy state (at 50\% each) to wherever
clear ferromagnetic behavior emerges. The accumulation of record events
grows logarithmic with time in the glassy regime, with a sharp
transition at a specific admixture into the ferromagnetic regime where
such activations saturate quickly. We show this effect both for the
Edwards-Anderson model on a cubic lattice as well as the Sherrington-Kirkpatrick
(mean-field) spin glass. While this transition coincides with a previously
observed zero-temperature equilibrium transition in the former, that
transition has not yet been described for the latter.
\end{abstract}
\maketitle

\section{Introduction\label{sec:Introduction}}

The morphology of complex energy landscapes~\cite{Frauenfelder96,Wales03},
and the parameters that control it, are of continuing interest in
a large variety of scientific endeavors, from protein folding and
evolutionary landscapes in biology~\cite{Kauffman1989,Frauenfelder1991,Stadler92,Bryngelson95,Lassig09},
the design of amorphous materials~\cite{Ball96,Wevers99,Schon00,Schon02,Schubert05,Aspelmeier06,Rylance06,Fischer08,Charbonneau13,Liao2019,ditBiot20},
to the hardness of combinatorial optimization problems~\cite{SA,Stadler92,mertens:00,Mezard06,Percus06,Salamon02}.
The challenges encountered in describing the geometry of the extremely
high-dimensional space of attainable configurations are enormous~\cite{Frauenfelder96,Wales03,Schon96,Sibani99,sg-equi2000,Sibani02,Salamon02,Schubert05,Heuer08}.
The structure of such energy landscapes hugely impacts the dynamics
of statistical systems evolving through them. While relaxation in
simple, smooth landscapes is rapid, like the exponential cooling of
a cup of coffee~\cite{Gould96}, relaxation in complex energy landscapes
can possess a myriad of metastable states to temporarily or permanently
trap any dynamic process. In turn, simple relaxation processes can
serve as diagnostic tools to explore features of landscapes~\cite{SA,Schon96,Sibani99,Sibani02,Dall03,BoSi,Bouchaud01}.
It is particularly enticing when it is possible to discover universal
aspects of such landscapes that allow to categorize those features
and, ultimately, predict and control dynamic behavior.

Variations in temperature can be used to take a full measure of landscapes.
At high temperature correspondingly higher echelons in energy get
explored, while annealing or quenching is used to trace out a descent
through the landscape towards configurations of lower energy. A conceptually
simple protocol consists of preparing a system at a high temperature,
where it equilibrates easily, and then instantaneously quenching it
down to a fixed, low temperature, to explore how it relaxes towards
equilibrium thereafter. Such an ``aging'' protocol~\cite{Struik78},
when applied to systems in a complex energy landscape, elicits quite
subtle relaxation behaviors which, unlike for the coffee mentioned
above, keeps the system far from a new equilibrium for very long times.
Anomalously slow relaxation and full aging in a complex landscape
ensues when downward paths are obstructed by barriers, energetic or
entropic, that trap the system in neighborhoods with many local minima.

The aging phenomenology is associated with memory effects by which
the current activity is imprinted by a dependence on the waiting time
$t_{w}$ since the quench. For a wide class of systems, generally
considered to be glassy, it is found that correlations, instead of
being time translational invariant, $G(t,t_{w})\sim f(t-t_{w})$,
roughly depend on a ratio, $G\sim f(t/t_{w})$. Although memory effects
in out-of-equilibrium systems are generally of interest, that fact
alone is not sufficient to categorize its energy landscape as complex
or glassy. To emphasize this fact, and to provide a deeper insight
into the relation between landscape morphology and aging dynamics,
we investigate here the aging in families of models that interpolate
between a well-known spin glass~\cite{Edwards75,Sherrington75} and
the corresponding ferromagnet~\cite{Manssen15}. Albeit glass and ferromagnet exhibit
similar scaling with age $t$, it stands to reason that the aging
dynamics of a homogeneous ferromagnet differs significantly from that
of a glass. In contrast to the hierarchical, multimodal energy landscape
of a glass~\cite{PSAA,Teitel85,Sibani86,Hoffmann88,Sibani89}, that
of a ferromagnet is smooth. Yet, much of the literature leads to the impression that
relaxation via coarsening in a ferromagnet and glassy aging are 
 synonymous~\cite{komori99,Mayer04,Biroli05}. Technically, one could argue
that the fact that in either extreme a growing length-scale emerges
is indicative of coarsening domains. We posit that the process by
which those length-scales grow with age, logarithmically in the glassy
case and with a power-law for a ferromagnet, is fundamentally different.

As discussed in Ref.~\cite{Shore92}, energy barriers that scale
with the size of a domain to be flipped imply that further growth
in those domains is curtailed to be merely logarithmic in time. Such
a feedback does not emerge in the coarsening of a ferromagnetic Ising
system, where energy barriers remain insensitive to the size of the
domain to be flipped. Accordingly, the landscape of a glassy system
has a \emph{hierarchical} structure in that, the lower an energy it
has reached, the higher the barriers get, and thus, the harder it
becomes to escape local minima~\cite{Robe16}. In a homogeneously
coarsening system, energy barriers remain largely independent of the
depth reached within the landscape, providing some roughness and metastability
but of bounded scale beyond which the structure is relatively smooth.
Within the aging process, this difference manifests itself dramatically
in the manner that the relaxing system responds to fluctuations, as
illustrated in Fig.~\ref{fig:Traces}. In a ferromagnet, average
fluctuations in energy, beyond some low, fixed threshold, suffice
to cross typical barriers, often followed by disproportionately large
expulsions of heat. In contrast, to advance glassy systems with diverging
energy barriers, mere average fluctuations become ineffective. To
be able to relax, those fluctuations have to produce ever new records in their size
to overcome ever steeper barriers, which is the basis of what is called Record 
Dynamics (RD)~\cite{Sibani03}. Such record-production, decorrelated
by a wide separation in time, is known to unfold only on a logarithmic
scale~\cite{SJ13,Sibani20}.

Although irrelevant for the cases studied in Ref.~\cite{Shore92}
(and here), it should be noted that entropic effects can become dominant
and may entail diverging free-energy barriers with domain size, even
in an otherwise homogenous system. One example is a 3-spin Ising ferromagnet~\cite{Mezard02}.
Systems driven by entropic barriers, such as the free volume in a
hard-core colloidal system, are referred to as ``structural'' glasses.
In those systems, a hierarchical free-energy landscape emerges dynamically~\cite{Liao2019}.

In the following, we define a simple coarse-graining procedure, counting
the number of ``valleys'' traversed in the energy landscape, that
effectively probes the impact of fluctuations on the aging dynamics.
Our focus on landscape morphology reveals the nature of the irreversible, intermittent events that
allow the expulsion of excess energy from the system. It shows a dynamical
transition between a glassy and a ferromagnetic relaxation regime
based on this measure. In that, we reproduce similar findings using scaling exponents 
of two-time correlation functions in the thermodynamic limit~\cite{Manssen15}, which
indicated that this dynamical transition is closely related with a zero-temperature equilibrium
transition between a glass and a ferromagnet~\cite{Hartmann1999}.
That this transition transcends into the non-equilibrium realm highlights the fact
 that aging in a glassy system is a distinct process from what is found 
 in homogeneous systems, characteristic of a distinct, hierarchical landscape.

Our paper is organized as follows: In the next Section~\ref{sec:Models},
we introduce the families of Ising spin models we employ in our study.
In Sec.~\ref{sec:Aging-and-Record}, we will discuss RD
and the measures we will apply to detect record fluctuations. In Sec.~\ref{sec:Numerical-Results},
we present the results of our investigation, and we conclude in Sec.~\ref{sec:Conclusion}.

\section{Models\label{sec:Models}}

Ising spin systems, consisting of spin variables $\sigma_{i}=\pm1$,
have been widely used, first of all as ferromagnets, to model spontaneous
symmetry breaking and continuous phase transitions~\cite{Plischke94}.
With the random admixture of anti-ferromagnetic bonds, they have also
served as models for disordered materials and glasses generally~\cite{Edwards75,Sherrington75,MPV,F+H}.
The relevance of such spin models reaches far beyond physics, into
biological and sociological applications, for example~\cite{Stein13}.
Here, we are employing families of such spin models that interpolate
between the randomly disordered spin glass on a cubic lattice, called
the Edwards-Anderson model (EA)~\cite{Edwards75}, as well as its
mean-field version, the Sherrington-Kirkpatrick model (SK)~\cite{Sherrington75},
on one side and the respective homogeneous ferromagnetic systems~\cite{Plischke94}
on the other. Each system consists of a random mixture of ferromagnetic
and anti-ferromagnetic bonds $J$ between neighboring spins $\sigma_{i}$
and $\sigma_{j}$ that are drawn from a distribution $P(J)$ we have
chosen to be bi-modal, i.e., $J_{ij}=\pm J_{0}$, with energy units
such that $J_{0}=1$ in 3D and $J_{0}=1/\sqrt{N}$ in the mean field
case. A fraction $p$ of ferromagnetic bonds is balanced out with
a fraction $1-p$ of anti-ferromagnetic bonds such that 
\begin{equation}
P(J)=p\delta\left(J-J_{0}\right)+(1-p)\delta\left(J+J_{0}\right).\label{eq:P(J)}
\end{equation}
For each, the Hamiltonian (without external field) reads 
\begin{equation}
H=-\sum_{\left\langle ij\right\rangle }J_{ij}\sigma_{i}\sigma_{j},\label{eq:Hamiltonian}
\end{equation}
where $\left\langle ij\right\rangle $ refers to all extant bonds
between neighboring spins $\sigma_{i}$ and $\sigma_{j}$, either
on a cubic lattice for EA or all mutual pairs of spins for SK.

In the family EA of models we study on the cubic lattice~\cite{Hartmann1999,Manssen15},
we change the admixture of bonds by varying $p$ between $\frac{1}{2}\leq p\leq1$, 
from the pure glass with an equal mix of bonds ($p=\frac{1}{2}$)
to a pure ferromagnet when all bonds are ferromagnetic ($p=1$). The
situation is more complicated for SK, where already a sub-extensive
excess of ferromagnetic bonds, away from the pure glass, results in
ferromagnetic behavior. Specifically, since all $N$ spins are mutually
connected, there are $\frac{1}{2}N(N-1)$ bonds, and it only takes
an imbalance between either type of bond, merely of order\textcolor{black}{{}
$\backsim\sqrt{N}$, to achieve ferromagnetic ordering. Thus, we define
a family of mean-field models parametrized by $\alpha$ with $p=\frac{1}{2}+\frac{\alpha}{\sqrt{N}}$,
varying between $0\leq\alpha\leq2$ to explore the full range} of
behaviors~\footnote{Note that in the bond matrix $J_{ij}$ of SK there are $\sim\frac{N^{2}}{2}$
bonds, thus, $\alpha$ skews an equal mix of ferromagnetic and anti-ferromagnetic
bonds with an imbalance of $O(N^{\frac{3}{2}})$, a vanishing fraction
of all bonds but significantly larger than random fluctuations between
bond types of $O(N)$. }.

\section{Aging and Record Dynamics\label{sec:Aging-and-Record}}

\subsection{Simulation of Quenches in Spin Glasses\label{subsec:Simulation-of-Quenches}}

The distinction between slow relaxation in glassy versus homogeneous
systems is succinctly analyzed in the simplest conceivable protocol
of a hard quench from an easily equilibrated high-temperature state
into an ordered phase, whether glassy or ferromagnetic, crossing a
phase transition in the process. Such a pure aging protocol has been
studied extensively in the last 40 years~\cite{Struik78,Hutchinson1995,Vincent96,SJ13,Biroli05}.
In this process, the system is thrown far out of equilibrium, left
with an enormous amount of excess heat to be released to the bath
to be able to descent deeper into its energy landscape to reach states
with the appropriate (equilibrium) energy.

To facilitate such a quench for the family of Ising spin models considered
in our study, for each instance at time $t=0$, we initiate with randomly
assigned spins, either $\sigma_{i}=\pm1$, which corresponds to $T=\infty$,
and run the simulation for $t>0$ at a low, finite temperature. For
our family of models on the cubic lattice, the critical temperature
for a transition into an ordered state varies from $T_{c}\approx1.1J_{0}$
in EA~\cite{BaityJesi2013}, to about $T_{c}\approx4.5J_{0}$ for
the ferromagnet~\cite{Butera2000}. In our Monte Carlo simulations,
we quench to $T_{q}=0.7J_{0}$ for all $p$, similar to Ref.~\cite{Manssen15},
and monitor the aging process for about $10^{5}$ sweeps. For the
family of mean field models, we only vary the admixture of ferromagnetic
bonds minutely, so that the transition temperature does not deviate
much from that of SK, which is known to be $T_{c}=J_{0}$~\cite{Sherrington75}.
Here, we also quench to $T_{q}=0.7J_{0}$ throughout. For each value
of $p$ in our study, we have averaged results over at least $10^{4}$
realizations.

\begin{figure}
\vspace{-0.5in}
\hfill{}\includegraphics[bb=0bp 40bp 750bp 720bp,clip,width=1\columnwidth]{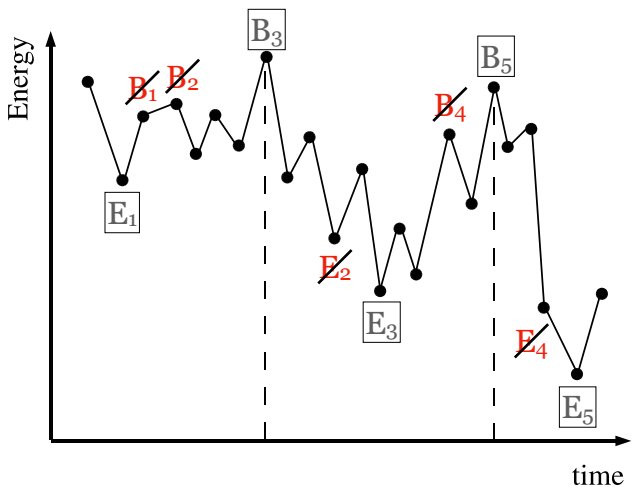}\hfill{}

\caption{\label{fig:DefValley}Illustration of the definition of valleys. The
trace through an energy landscape produces a time sequence of energy
records (${\rm E}_i$) and of barrier records (${\rm B}_j$), relative to the most recent
``${\rm E}_i$''~\cite{BoSi,Dall03}. Only the highest and lowest records
of the ``${\rm E}_i$" and ``${\rm B}_j$'' are kept to give a strictly alternating sequence
``$\ldots{\rm E}_{i_0}{\rm B}_{j_1}{\rm E}_{i_1}{\rm B}_{j_2}{\rm E}_{i_2}\ldots$''. Then, any 
sequence ``${\rm B}_{j_1}{\rm E}_{i_1}{\rm B}_{j_2}$'' demarcates a valley (vertical
lines).}
\end{figure}

\subsection{Valleys in an Energy Landscape\label{subsec:Valleys}}

Key to our analysis of coarsening versus glassy relaxation is the
definition of a measure that can serve to distinguish the effect of
fluctuations on the irreversible events by which a system relaxes.
One such measure has been provided by Dall and Sibani in Ref.~\cite{Dall03}.
There, the internal energy of an entire system of finite size is monitored
to observe its time-trace for the ensuing quench. Since the system
is expelling energy into the bath to relax, on average, the energy
gradually decreases, albeit via localized, intermittent events~\cite{Sibani05},
in line with experimental observations of glassy systems~\cite{Bissig03,Cipelletti03a,Mayer04,Yunker09}.
In particular, record-sized fluctuations are needed for a glassy system
to relax, according to RD~\cite{Sibani03,Robe16}.

As illustrated in Fig.~\ref{fig:DefValley}, Dall and Sibani defined
the ``valley'' production as an observable.~\cite{Dall03} in the
following way: Let ${\rm E}$ be the up-to-now lowest energy value encountered
up to time $t$ and let $E(t)$ be the instantaneous energy. In turn,
let the ``barrier'' ${\rm B}$ be the up-to-now highest energy attained,
\emph{relative} to the most recent E, i.e., ${\rm B}=E(t)-{\rm E}$.
An energy trace then maps into a random sequence of symbols, like in Fig.~\ref{fig:DefValley},
$\ldots{\rm E}_1{\rm B}_1{\rm B}_2{\rm B}_3{\rm E}_2{\rm E}_3{\rm B}_4{\rm B}_5{\rm E}_4{\rm E}_5\ldots$. 
Note that the trace can generate a sub-sequence
of records in the lowest energy, i.e., multiple E's in a row, before
it encounters its next higher (record) barrier B, and also a sub-sequence of
such B's before it meets the next E, and so on. Clearly, it is the
latest E or B in either type of sub-sequence that is significant:
Each prior one is merely transitory, while the last one supersedes
each prior one as record that reaches its ultimate significance \emph{only
after} a new record fluctuation in the opposite direction is attained.
Thus, we squash the entire sequence into a strict alternation between
E and B, as the stricken letters in Fig.~\ref{fig:DefValley} imply, which then yield: $\ldots{\rm E}_1{\rm B}_3{\rm E}_3{\rm B}_5{\rm E}_5\ldots$.
Then, a ``valley'' is defined as the part of the trace between two
consecutive record barrier-crossings, as indicated by vertical dashed
lines there. If the ground state were to be reached, i.e., no further energy 
minimum could be found,  the sequence would terminate, of course.

To focus truly on locally correlated record barrier crossings, it
would be useful to refine this definition of valley~\cite{Sibani18}.
However, unless a system gets too large, with too many simultaneous
but spatially distant quakes, by considering a small enough system
these events become sufficiently rare to dominate the fluctuations
in the entire system trace, instead of being ``washed out'' by overlapping ones. This
point illustrates also that, to understand a thermodynamic system
\emph{out-of-equilibrium}, it is often \emph{not} helpful to take
the thermodynamic limit.

\begin{figure}
\hfill{}\includegraphics[bb=30bp 275bp 450bp 580bp,clip,width=1\columnwidth]{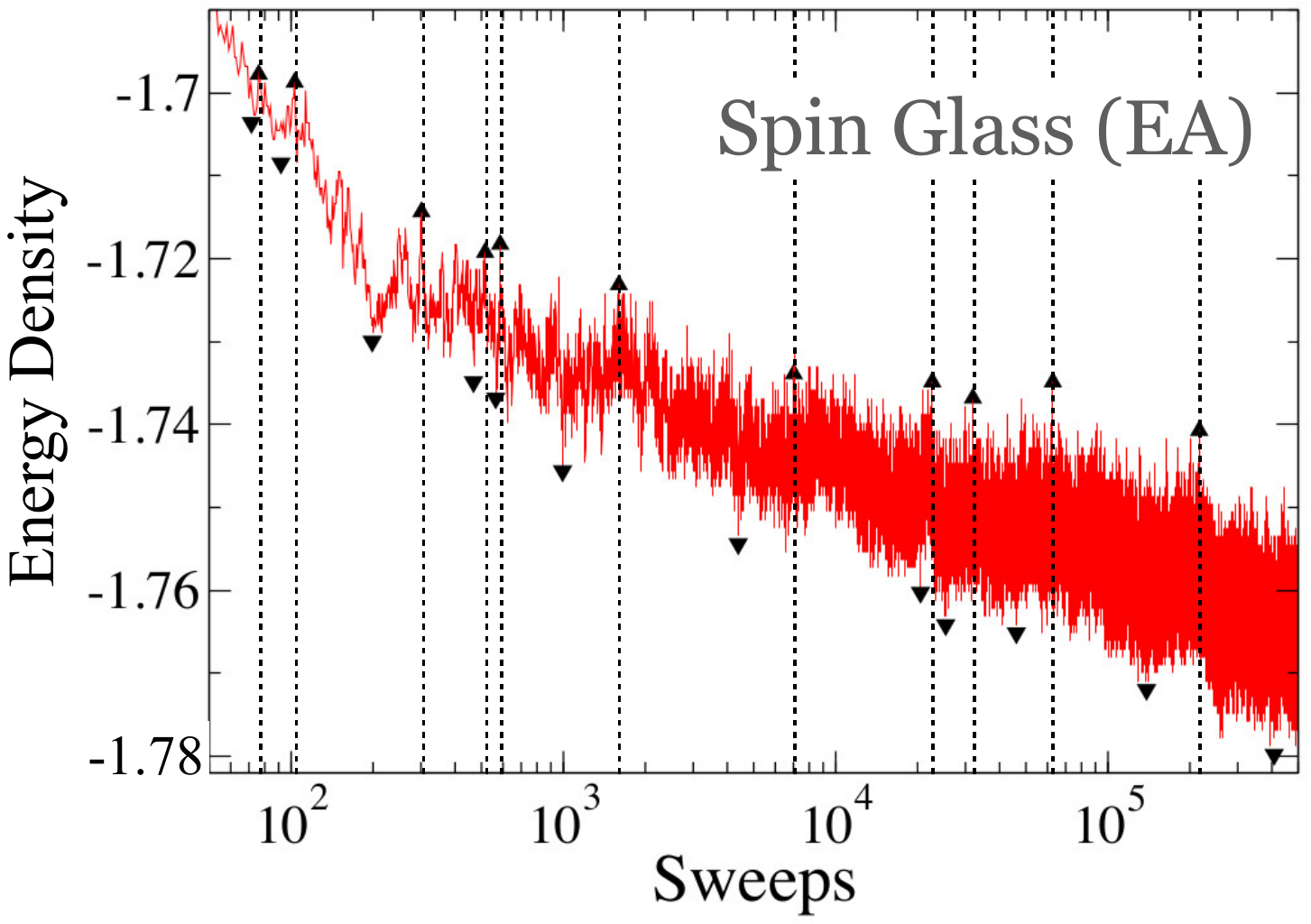}\hfill{}
\vspace{-0.6in}

\hfill{}\includegraphics[bb=30bp 275bp 450bp 575bp,clip,width=1\columnwidth]{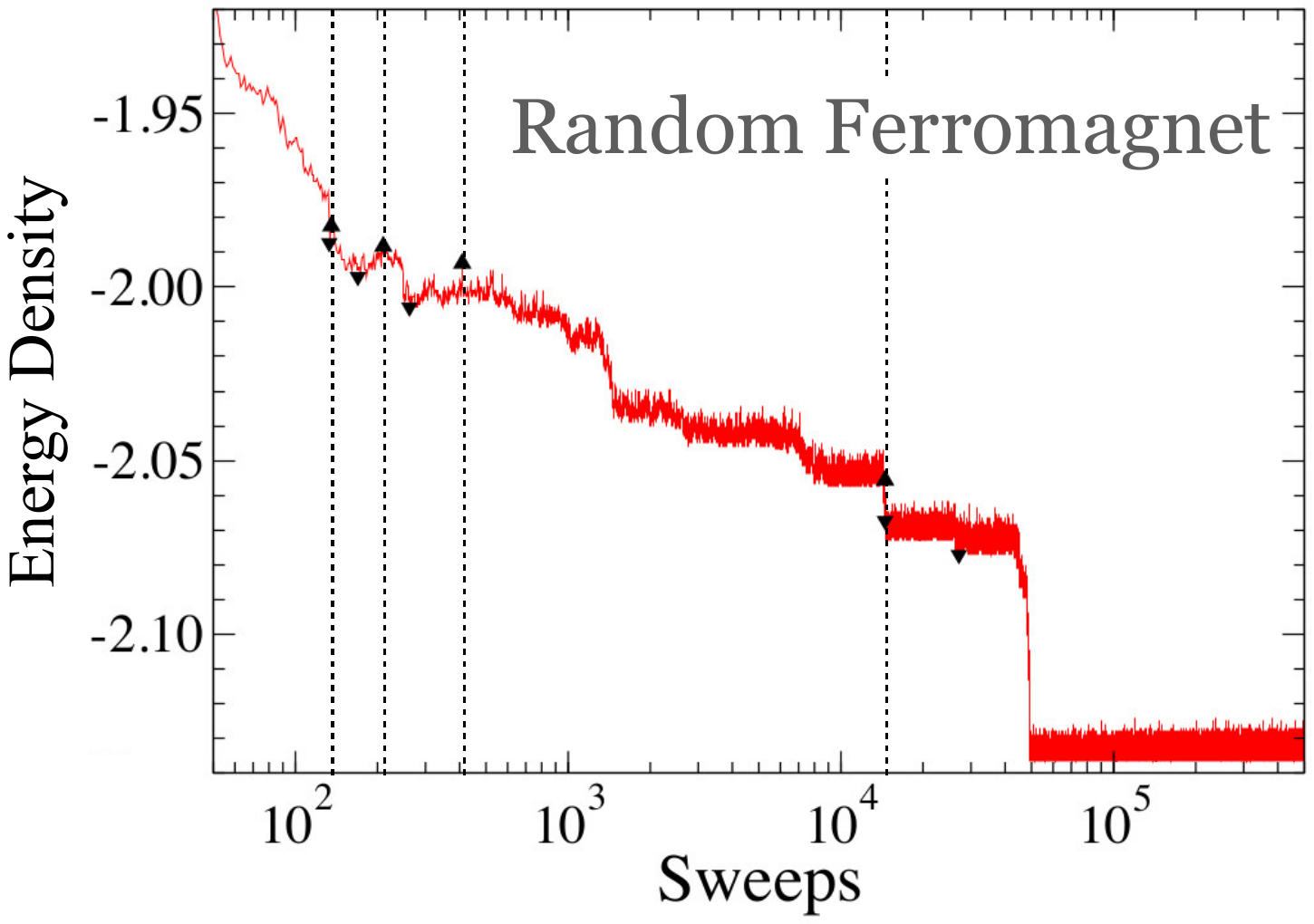}\hfill{}
\vspace{-0.3in}

\caption{\label{fig:Traces}Typical trajectories of an aging process through
the energy landscape of the spin glass model on a \emph{3d}-lattice
with $L^{3}=16^{3}$ spins and a fraction $p$ of ferromagnetic bonds
and $1-p$ anti-ferromagnetic bonds, here with $p=0.5$ (top) and
$p=0.85$ (bottom). Energy ($\blacktriangledown$) and barrier ($\blacktriangle$)
records, as defined in Fig.~\ref{fig:DefValley}, are marked along
each trajectory, where the vertical dashed lines indicate the transition
between consecutive valleys. While the energy decreases, on average,
gradually as a logarithm in time with an ongoing but random production
of further records in the glassy case ($p=0.5$), the more ferromagnetic
system ($p=0.85$) expels energy in a few large events which appear
to be triggered by typical fluctuations, record-sized fluctuations
are seemingly irrelevant.}
\end{figure}

Examples of a valley sequence from our simulations is shown for single
energy traces in Fig.~\ref{fig:Traces} for the EA spin glass (top)
and the corresponding ferromagnetic system (bottom) on a cubic lattice.
These plots exemplify the stark difference in the effect of fluctuations
on either type of system that we discuss in the following.

\subsection{Dynamics Driven by Record-Sized Fluctuations\label{subsec:Dynamics-Driven-byRD}}

As alluded to in the introduction, glassy and otherwise homogeneous
systems such as a ferromagnet distinguish themselves in the manner
fluctuations affect their relaxation dynamics. In the latter, barriers
are comparably low and remain invariant independent of the depth within
the landscape and, thus, of the age of the process. As Fig. \ref{fig:Traces}
exemplifies, large releases of energy are preceded by typical fluctuations
at any stage of the process. Fewer events, like the evaporation of
a domain in coarsening, happen not because individual events become
so much harder but rather because so many fewer events can happen
when only few domains is left. Larger domains may take a little more
time to evaporate, as meandering interfaces need to find each other
and collide, but such an entropy barrier does not dominate the otherwise
domain-size independent energetic barriers~\cite{Shore92}. Yet,
ordinary fluctuations suffice to bring those interfaces together.

In the glassy system, however, it is the barrier height growing with
domain size that decelerates the event-rate. Although many domains
remain available even after a long aging time, few muster the chance
fluctuation required to break up. In a landscape with those barriers,
ordinary fluctuations become ineffective to drive the relaxation process.
They merely ``rattle'' the system during increasingly longer quasi-equilibrium
interludes. Only rare, extraordinary large, in fact, record-size fluctuations
manage to scale such barriers to expel excess heat, advance the relaxation,
and grow domain size, minutely.

These features, widely shared across many disordered materials, have inspired
the phenomenological description at the basis of RD~\cite{Sibani03}. The definition of valleys 
in the preceding section provides an especially adapt measure for this phenomenology. 
In RD, the relaxation process of a non-equilibrium system after a
hard quench is determined by large (i.e., record-sized), irreversible fluctuations which
move the system from one meta-stable state to the next (usually only
marginally more stable than the last one) within its complex energy
landscape~\cite{Becker14,Robe16}. This can be thought of as the
system overcoming energy barriers in a hierarchical energy landscape~\cite{PSAA,Teitel85,Sibani86,Hoffmann88,Sibani89}.
The rate $\lambda(t)$ of such record events, also termed ``quakes'',
decelerates with time as $1/t$. Hence, the expected number of
events in a time interval {[}$\mathit{t}$, $t_{w}${]}, is 
\begin{equation}
\left\langle n(t,t_{w})\right\rangle \wasypropto\intop_{t_{w}}^{t}\lambda\left(t^{\prime}\right)dt^{\prime}\wasypropto\mathrm{ln}\left(\frac{t}{t_{w}}\right)\label{eq:lograte}
\end{equation}
implying that the dynamics of the system is self-similar in the logarithm
of time. That time-homogeneity is a common feature of many aging 
systems~\cite{Hutchinson1995,Biroli05,PhysRevLett.89.217201}.
In our studies here, we are more concerned with the rate of events
$\lambda(t)$ and the logarithmic growth of observables in time. The
dependence on waiting time $t_{w}$ has been the focus elsewhere~\cite{Sibani05,Becker14,Robe16}.

\begin{figure}
\hfill{}\includegraphics[bb=0bp 10bp 742.5bp 540bp,clip,width=1\columnwidth]{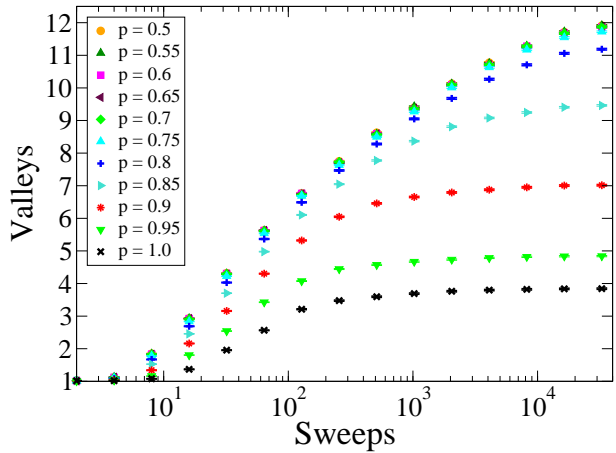}\hfill{}
\vspace{-0.3in}

\caption{\label{fig:Valley}Average number of valleys in EA, as defined in Fig.~\ref{fig:DefValley},
that are traversed with time after a quench to $T=0.7J_{0}$ in a
$L^{d}=16^{3}$ spin glass with a fraction $p$ of ferromagnetic bonds
and $1-p$ anti-ferromagnetic bonds. For $p\protect\leq0.75$, the
generation of valleys evolves essentially independent of $p$, while
for a larger admixture of ferromagnetic bonds valley generation progresses
to cease ever more rapidly and the number of valleys reached plateaus.}
\end{figure}

\begin{figure}
\hfill{}\includegraphics[bb=0bp 10bp 742.5bp 540bp,clip,width=1\columnwidth]{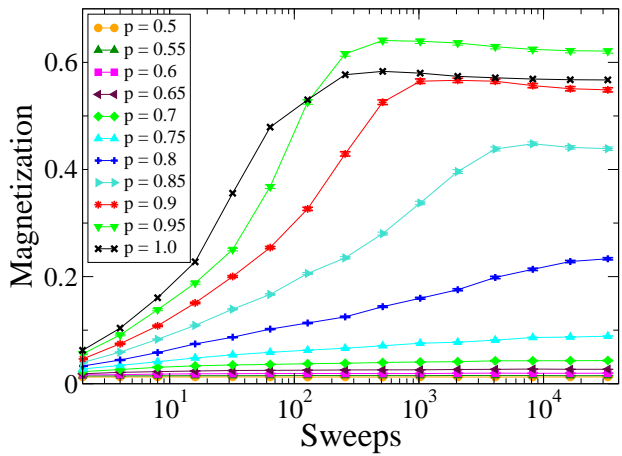}\hfill{}
\vspace{-0.3in}

\caption{\label{fig:magnetization}Average magnetization per spin in EA, $\left\langle m\right\rangle $,
observed with time after a quench during the ensuing aging process,
as described in Fig.~\ref{fig:Valley}. Like there, systems with
$p\protect\leq0.75$ behave glassy in a $p$-independent manner with
little discernible magnetic ordering, while the more ferromagnetic
systems become increasingly more ordered.}
\end{figure}

\begin{figure}
\hfill{}\includegraphics[bb=0bp 10bp 742.5bp 540bp,clip,width=1\columnwidth]{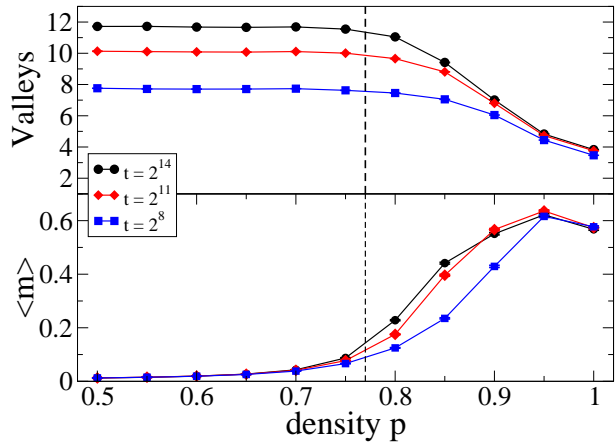}\hfill{}
\vspace{-0.3in}

\caption{\label{fig:snapshopt}Finite-time snapshots for EA of the numbers of valleys
generated (top) and the corresponding magnetization per spin, $\left\langle m\right\rangle $
(bottom), as a function of ferromagnetic bond fraction $p$ for three
different times, taken from the data at $T=0.7J_{0}$ shown in Fig.~\ref{fig:Valley}
and Fig.~\ref{fig:magnetization}, respectively. The vertical line
at $p_{c}=0.77$ indicates the zero-temperature transition found in
Ref.~\cite{Hartmann1999} between a glassy and a ferromagnetic phase.}
\end{figure}

\begin{figure}
\hfill{}\includegraphics[bb=0bp 10bp 742.5bp 540bp,clip,width=1\columnwidth]{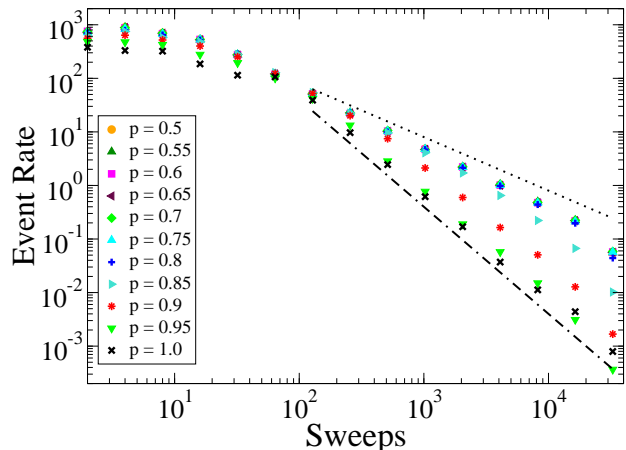}\hfill{}
\vspace{-0.3in}

\caption{\label{fig:eventrate}Instantaneous rate of record barrier crossing
events in EA, as defined in Fig.~\ref{fig:DefValley}, with time after
the quench, as described in Fig.~\ref{fig:Valley}. Asymptotically,
for larger times, that rate varies as a power-law with a seemingly
hyperbolic decline, $\sim1/t$ (dotted line), for all $p<0.75$ to
an almost quadratic decline, $\sim1/t^{2}$ (dash-dotted line), for
larger $p$.}
\end{figure}

\section{Numerical Results\label{sec:Numerical-Results}}

\subsection{Edwards-Anderson Model\label{subsec:Edwards-Anderson-Model}}

Applying the measure of a valley number defined in Sec.~\ref{subsec:Valleys}
to the cubic Ising spin model introduced in Sec.~\ref{sec:Models}
provides a notable distinction between glassy and homogeneous coarsening
behavior, as Fig.~\ref{fig:Valley} shows. For all $p<p_{c}\approx0.77$,
the critical threshold found in Ref.~\cite{Hartmann1999}, we find
that the valley count progresses logarithmically in time (in fact,
like the root of that logarithm~\cite{Becker14}), consistent with
Eq.~(\ref{eq:lograte}). For larger values of $p$, the valley count
slows ever more significantly to eventually plateau at a finite value,
apparently. All the results shown here were obtained for systems with
$N=16^{3}=8096$ spins, using periodic boundaries, since we found
very little variation with system size for larger $N$.

The fact that the underlying ordered state is either glassy or ferromagnetic
affords us to also measure the increase in magnetization with time,
as demonstrated in Fig.~\ref{fig:magnetization}. This measure actually
exhibits a more dramatic transition between the glassy and the ferromagnetic
case, as consecutive snapshots of both, the valley count as well as
the magnetization, are shown in Fig.~\ref{fig:snapshopt} for a progression
of times that increases by a factor of 8. In these plots, we have
also marked the zero-temperature transition at $p_{c}\approx0.77$,
which proves consistent asymptotically with the transition out of
the glassy relaxation behavior.

Finally, we can also look at the instantaneous rate of barrier crossing
events, effectively the derivative of the valley production, i.e,
inverting the integral in Eq.~(\ref{eq:lograte}). Indeed, throughout
the glassy regime, the rate decelerates roughly hyperbolically, in
accordance with the RD predictions. {[}Note that this could miss a
minor logarithmic correction, such as $\lambda(t)\sim1/(t\sqrt{\ln t})$,
for instance, needed to get $\sqrt{\ln t}$ for the valley production
in Fig.~\ref{fig:Valley}.{]} For $p>p_{c}$, in the ferromagnetic
coarsening regime, we notice that the rate falls off increasingly
sharper, ultimately about as $\sim1/t^{2}$. Consequently, its integral
stalls out into the plateaus seen in Fig.~\ref{fig:Valley}. Apparently,
domain mergers occur more rapidly, on a power-law scale, in coarsening
ferromagnets. Despite the rapid drop in the event rate, the average
domain size manages to increase as a power-law~\cite{Shore92}, because
later mergers expel larger amounts of excess heat, see Fig. \ref{fig:Traces}.
In case of the glass, each event expels on average a fixed amount
of heat, roughly. Therefore, both valley production and domain growth
proceed similarly (logarithmically), as an integral of the event rate,
since each activation has the same impact.

\begin{figure}
\hfill{}\includegraphics[width=1\columnwidth]{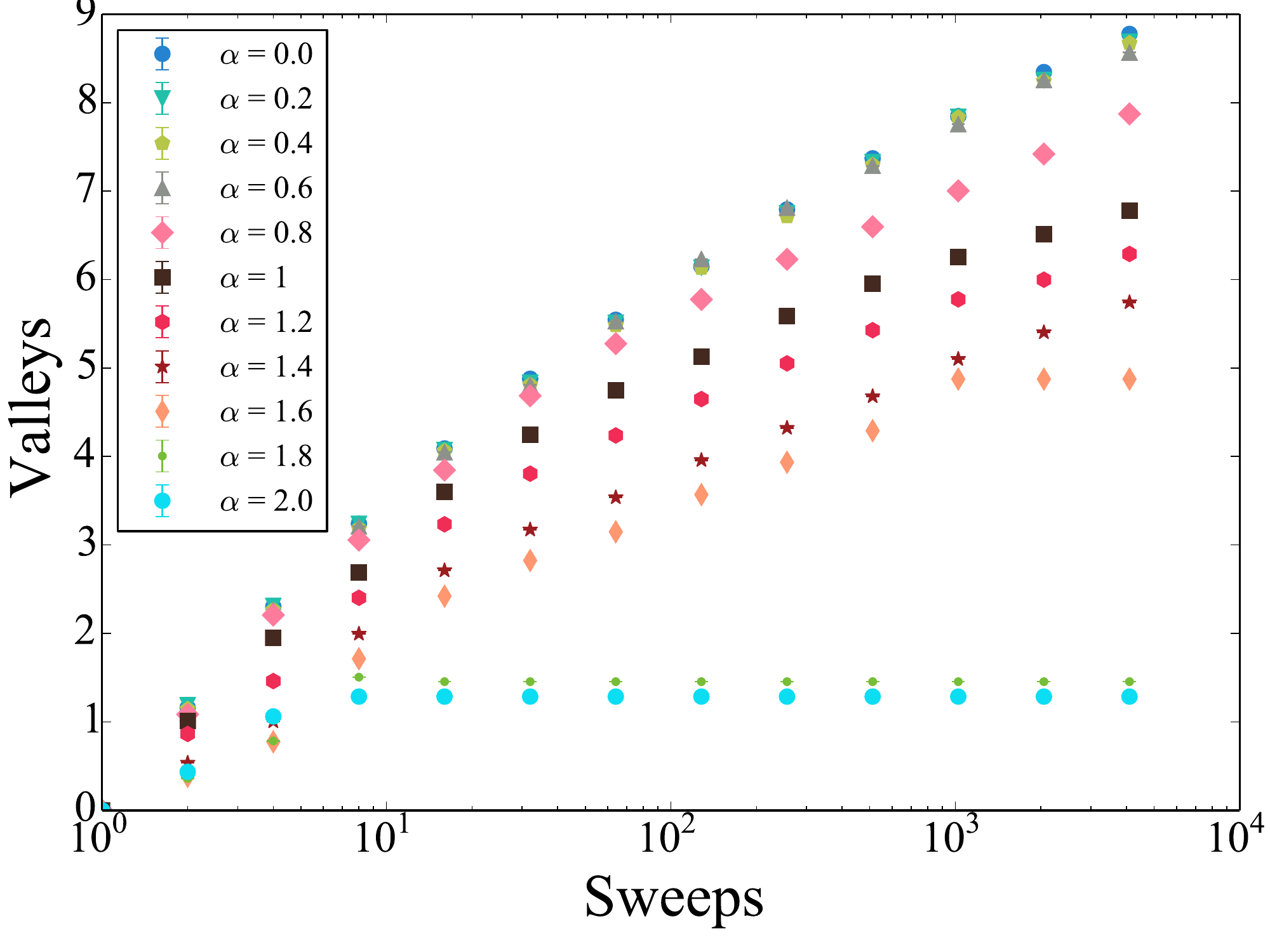}\hfill{}
\vspace{-0.3in}

\caption{\label{fig:valleySK}Number of valleys traversed during relaxation
ensuing after a quench of SK for different bond fractions $\alpha$
from a high temperature $T=\infty$ to $T=0.7J_{0}$, averaged over
an ensemble of trajectories for $N=2048$ spins. In the range $0.0\protect\leq\alpha\protect\leq0.6,$
the number of valleys traversed grows logarithmically and largely
independent of $\alpha$, indicating that the regime is glassy.}
\end{figure}

\begin{figure}
\hfill{}\includegraphics[width=1\columnwidth]{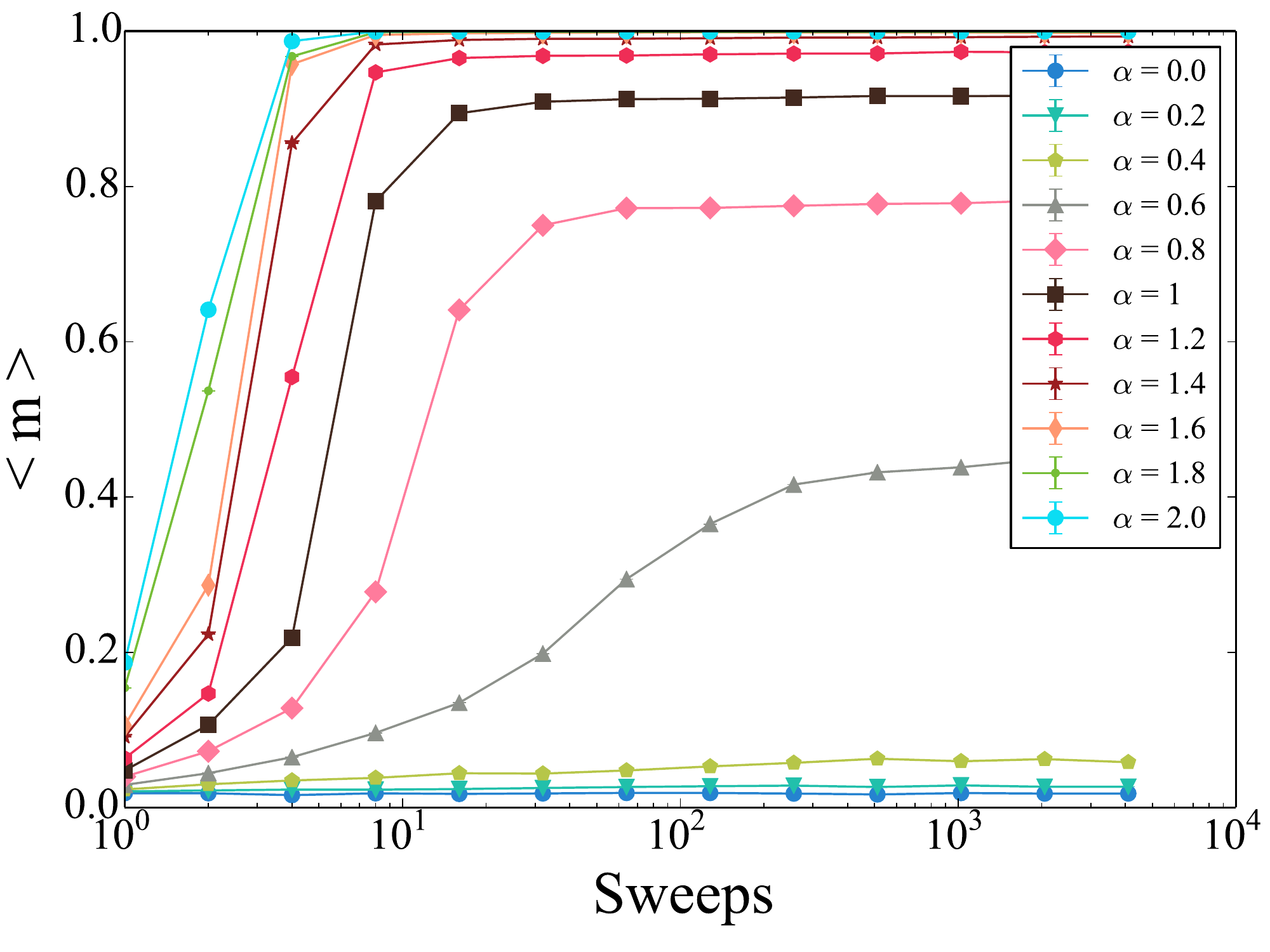}\hfill{}
\vspace{-0.3in}

\caption{\label{fig:MagnetizationSK}Average magnetization for SK in the same simulations
shown in Fig. \ref{fig:valleySK}. According to this measurement,
the system begins to order at $\alpha_{c}\approx0.6$, since a non-zero
magnetization in the long-time limit indicates that majority of the
spins have ferromagnetically ordered. The transition in magnetization
shown here is far more dramatic than in the valley counts, but nevertheless
affirms the same critical threshold.}
\end{figure}

\begin{figure}
\hfill{}\includegraphics[width=1\columnwidth]{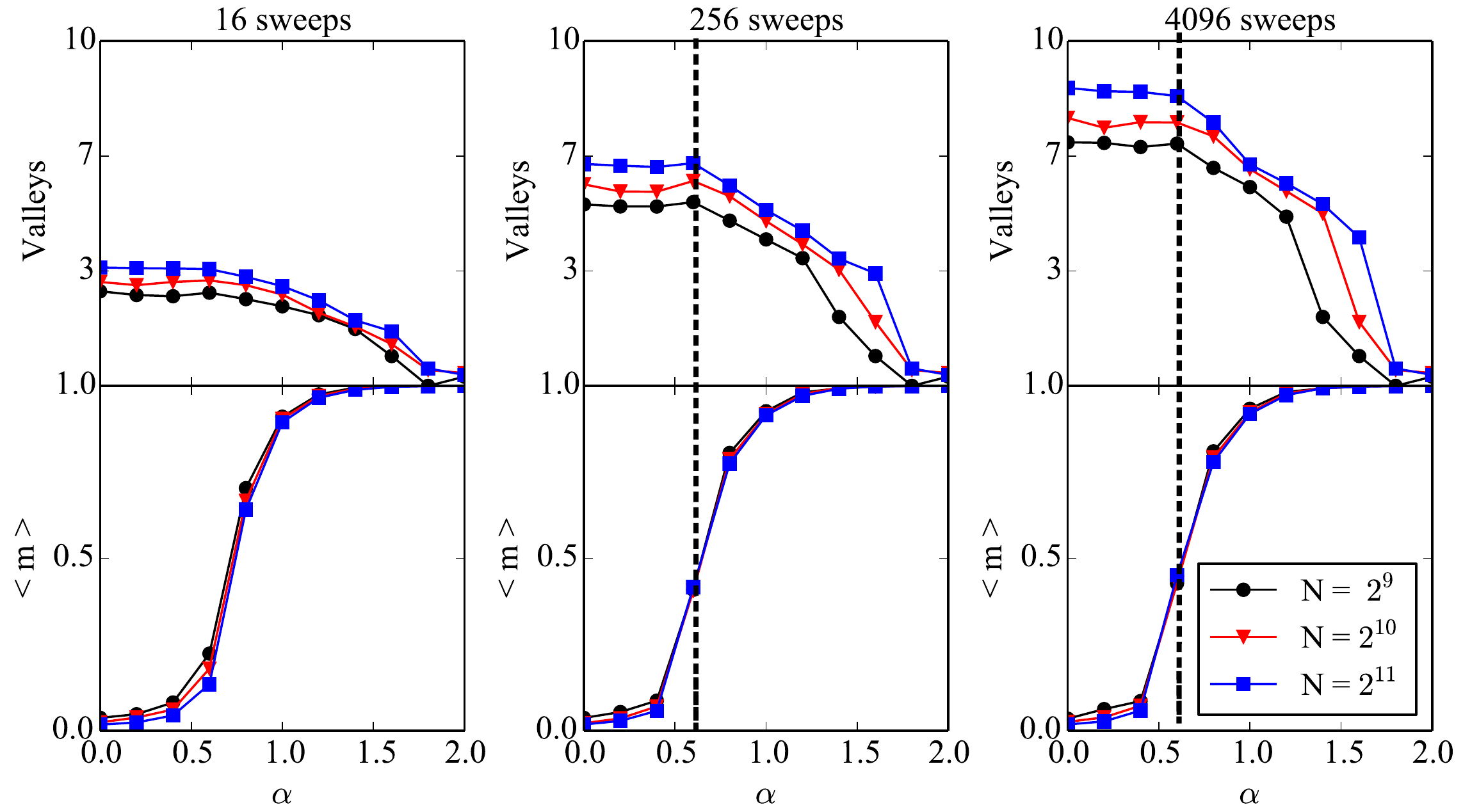}\hfill{}
\vspace{-0.3in}

\caption{\label{fig:Instant-avg}Instantaneous average valley counts and magnetization in SK
as function of $\alpha$ at different sweep-times $t=16$, 256 and
4096 from left to right, each for three different system sizes indicated
on the legend. The first row shows the average number of valleys,
and the second row shows the average magnetization. According to this
data, the valley production is time dependent as the sharpness of
the transition becomes more pronounced in the later sweeps. In contrast,
the magnetization appears to be saturated already early on, predicting
the critical threshold within 16 sweeps. Additionally, we see no system
size effects when using $\alpha$ as the parameter.}
\end{figure}

\subsection{Sherrington-Kirkpatrick Model\label{subsec:Sherrington-Kirkpatrick-Model}}

Using the valley counts defined in Sec. \ref{subsec:Valleys} as an
order parameter, we find a clear transition from a glassy regime to
a ferromagnetic one in the mean field as well. However, unlike for
EA on a cubic lattice, extending the neighborhood of each spin to
all others in the case of SK changes the dynamics, and we have to
explore the critical threshold at which the spin glass to ferromagnetic
transition takes place on a different scale. Mutual connections between
all spins require the number of ferromagnetic bonds to only slightly
exceed the number of antiferromagnetic bonds, in order to tip the
system into becoming ordered. The transition to the ferromagnetic
regime occurs almost immediate beyond a bond density of $p=0.5$,
with a strong system size dependence, forcing us to adapt a different
scale to observe it. To properly describe the behavior of SK, we therefore
reparametrize the bond density in terms of $\alpha$ via $p=\frac{1}{2}+\frac{\alpha}{\sqrt{N}}$.
Then, within the range of $0\leq\alpha\leq2.0$, we can localize a
transition that varies only slowly with size.

Similar to Fig. \ref{fig:Valley} for EA, in Fig. \ref{fig:valleySK}
we show the numbers of valleys found in a SK system with $N=2048$
spins. There appears to be a critical threshold at $\alpha_{c}\approx$
0.6. For $\alpha\leq0.6,$ the valley production increases about as
log$(t)$, essentially uniform with bond density, given the nearly
perfect overlap in the data. This is no longer case when $\alpha>0.6$,
where the production of valleys decreases gradually before plateauing
completely. While domains in the sense of geometric regions of a certain
length do not exist in a mean field system with long-range interactions,
individual spins develop clusters of increasingly ordered local fields
with some of their neighbors that entrench the system into deeper
valleys. It becomes increasingly more difficult for the system to
overcome the energy barrier of flipping the entire cluster, causing
the relaxation process to evolve logarithmically~\cite{Shore92}. 

That said, evidence of a critical threshold suggests that beyond $\alpha_{c}$,
the system changes its landscape dramatically. It exhibits an inclination
to order rapidly, facilitated by the fact that local fields of individual
spins immediately affect all others, as the evolution of magnetization
in Fig. \ref{fig:MagnetizationSK} suggests. Flat interfaces between
such clusters, as they may exist between domains in low-dimensional
lattices like EA, are absent here and any imbalance in size quickly
erodes inferior clusters. Therefore, despite the quantitative differences
pertaining to local structure between the Edwards-Anderson and Sherrington-Kirkpatrick
spin glass, our results suggest that the glassy behavior in both can
be attributed to the hierarchical nature of the energy landscape,
and the lack of it beyond the transition to ferromagnetic order, seen
both in Fig. \ref{fig:magnetization} and Fig. \ref{fig:MagnetizationSK}.

\begin{figure}
\hfill{}\includegraphics[width=1\columnwidth]{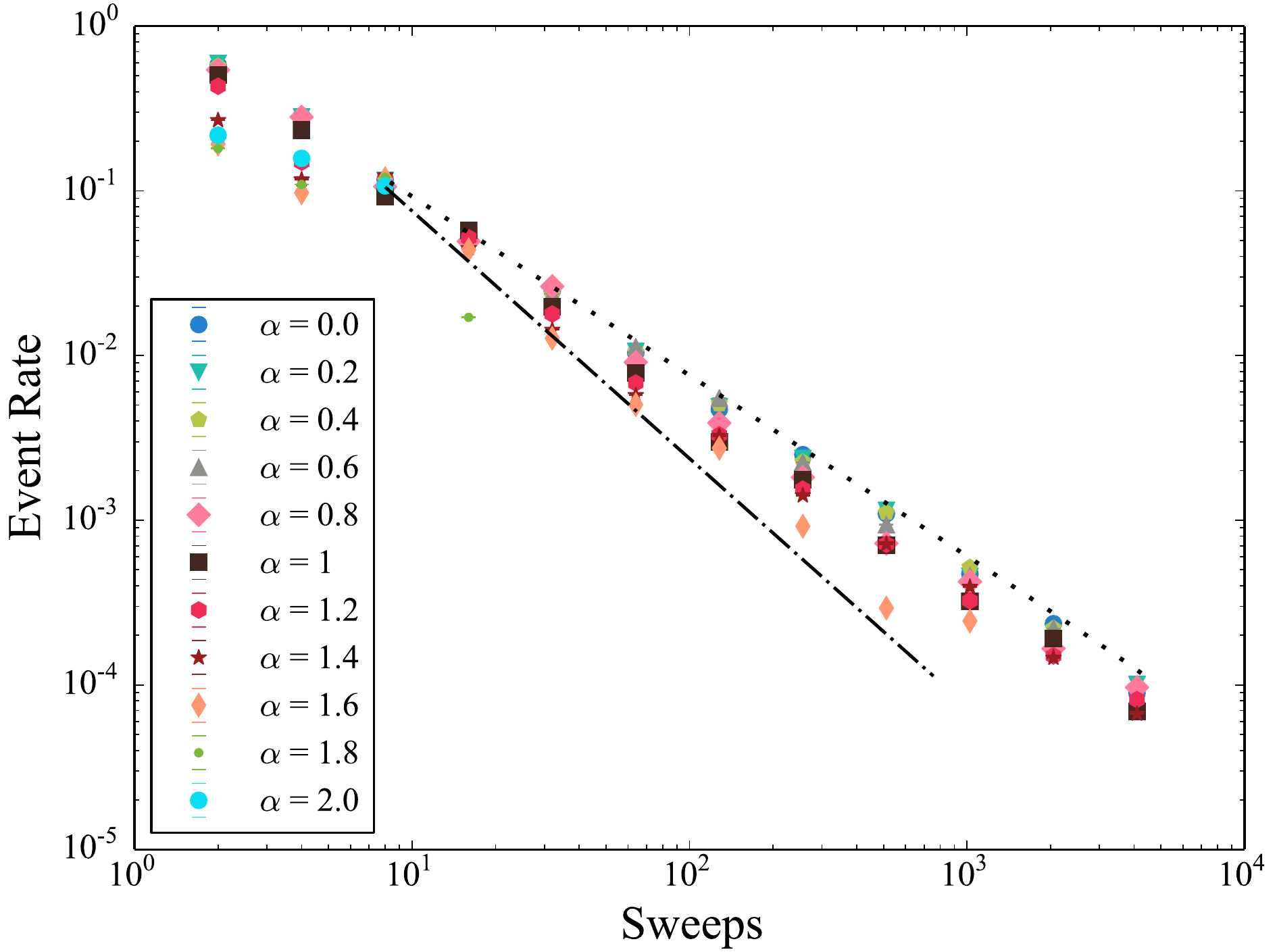}\hfill{}
\vspace{-0.3in}

\caption{\label{fig:eventrates-sk}Instantaneous rates for the number of record
barrier crossings as a function of time, for every $\alpha$-value in SK.
The instantaneous rate decreases as a power-law for all but the highest
admixture values. In the glassy regime, the decelerations is essentially
hyperbolic (dotted line), while the rate drops more sharply for $\alpha>0.6$, up
to roughly $t^{-1.5}$ at $\alpha=1.6$ (dash-dotted line), beyond which further record
events become immeasurably rare.}
\end{figure}

We have also checked the evolution of valley counts across different
system sizes and found only a minimal dependence of the transition
on larger size, as shown in Fig. \ref{fig:Instant-avg}. While the
relationship (or lack thereof) between the number of valleys encountered
and the bond admixture exhibits time dependence, the critical threshold
with regard to ordering already emerges after about two hundred sweeps.
There is clearly an agreement between valley statistics and the ferromagnetic
order parameter in suggesting $\alpha_{c}\thickapprox0.6$ as the
critical threshold.

Lastly, we look at the deceleration of the rate of record barrier
crossing events in Fig. \ref{fig:eventrates-sk}. As shown in Fig.
\ref{fig:eventrate} for the Edwards-Anderson model, the rate decays
with a power of time $t$. While there is a steeper deceleration in
the barrier crossing events for larger $\alpha$-values, the difference
between the exponents is quite subtle on this time scale within our
simulations. In the glassy regime, $\alpha<\alpha_{c}\approx0.6$,
the rate clearly decays hyperbolically, whereas it falls off steeper
above $\alpha_{c}$. However, for values $\alpha>1.6$, the fall-off
becomes so significant that new valleys are not encountered beyond
the first $\sim100$ sweeps.

\section{Conclusion\label{sec:Conclusion}}

Our study explores the distinction between glassy relaxation and ordinary
coarsening, which is often ignored in the description and analysis
of aging systems. Focusing on families of models that interpolate
between either extreme, we not only apply measures~\cite{Dall03,BoSi}
that clearly indicate the difference but also show a rather sharp
transition in the non-equilibrium behavior between those extremes
that, for the Edwards-Anderson model on a cubic lattice, appears to
coincide with the (equilibrium) zero-temperature transition between
spin glass and ferromagnet~\cite{Hartmann1999}. The corresponding
transition we find at a sub-extensive scale in SK seems to have been
unnoticed. 

While the distinction we are making between a coarsening (ferromagnetic)
and an aging (glassy) regime can be seen as semantic, considering
that both, algebraic as well as logarithmic growing domains, are commonly
portrayed as coarsening~\cite{Shore92}, the difference in dynamic
behavior after a quench is profound. The picture that emerges is one
of a largely convex landscape on one side with invariant energetic
barriers in the case of coarsening, a system that despite its often
complex network of fractal interfaces locally homogenizes rather quickly.
On the other side, we find a  hierarchical landscape~\cite{PSAA,Teitel85,Sibani86,Hoffmann88,Sibani89}
with energetic (and potentially entropic) barriers that grow with
deeper entrenchment within the landscape, rendering all but record
fluctuations ineffective for relaxation. 

 \bibliographystyle{apsrev4-1}
\bibliography{/Users/sboettc/Boettcher}

\end{document}